\documentclass[manuscript, nonacm]{acmart}

\AtBeginDocument{%
  \providecommand\BibTeX{{%
    \normalfont B\kern-0.5em{\scshape i\kern-0.25em b}\kern-0.8em\TeX}}}

\setcopyright{none}

\renewcommand{\b}[1]{\textbf{#1}}

\newcommand{\texting}[1]{\textit{messaging}}
\newcommand{\GoogleGlass}[0]{\textit{GG}}
\newcommand{\GlassMessaging}[0]{\textit{GM}}

\newcommand{\voicecommand}[1]{`\MakeUppercase{#1}'}
\newcommand{\buttoncommand}[1]{`{#1}'}
\newcommand{\NAME}[1]{\textless{}NAME\textgreater{}}

\renewcommand{\quote}[1]{``#1''}

\usepackage{subcaption}
\usepackage{multirow}
\usepackage{xcolor}
\usepackage{enumitem}
\usepackage{xspace}
\usepackage{booktabs}

\usepackage{sidecap} 
\sidecaptionvpos{figure}{hptb}

\usepackage[nomessages]{fp}
\usepackage{soul} 

\newlength\maxlen

\def\databarlength{xx.xx} 
\newcommand{\GlassTexting}[0]{\textit{GlassMessaging}}

\begin{document}

\title[\GlassTexting{}]{\GlassTexting{}: Supporting Messaging Needs During Daily Activities Using OST-HMDs}

\author{Nuwan Janaka}
\orcid{0000-0003-2983-6808}


\author{Jie Gao}
\orcid{0000-0001-5992-6471}


\author{Lin Zhu}
\orcid{0000-0003-0353-3534}


\author{Shengdong Zhao}
\orcid{0000-0001-7971-3107}

\author{Lan Lyu}
\orcid{0000-0001-5723-3575}

\author{Peisen Xu}
\orcid{0000-0003-1312-3061}


\author{Maximilian Nabokow}
\orcid{0000-0002-3417-4020}


\author{Silang Wang}
\orcid{0000-0002-1265-2220}

\author{Yanch Ong}
\orcid{0000-0001-6031-6064}

\affiliation{%
  \institution{National University of Singapore}
  \country{}
}

\renewcommand{\shortauthors}{Janaka et al.}

\begin{abstract}
The act of communicating with others during routine daily tasks is both common and intuitive for individuals. However, the hands- and eyes-engaged nature of present digital messaging applications makes it difficult to message someone amidst such activities. We introduce \GlassTexting{}, a messaging application designed for Optical See-Through Head-Mounted Displays (OST-HMDs). It facilitates messaging through both voice and manual inputs, catering to situations where hands and eyes are preoccupied. 
\GlassTexting{} was iteratively developed through a formative study identifying current messaging behaviors and challenges in common multitasking with messaging scenarios. 

\end{abstract}

\maketitle

\section{Introduction and Related Work}

The proliferation of mobile devices has transformed our means of communication, making applications (henceforth referred to as \textit{apps}) like WhatsApp, Telegram, Messenger, and WeChat commonplace \cite{curry_most_2022}. However, using these apps during daily tasks, such as cooking or walking, is hindered by their design, which demands extensive visual and manual interaction. Research reveals that individuals often use messaging apps while multitasking, with 13\% of messages sent on the move \cite{battestini_large_2010}.

Given this context, we question, \textit{``How can we refine mobile messaging for effective communication during routine multitasking?''} This brings us to Optical See-Through Head-Mounted Displays (OST-HMDs or Augmented Reality Smart Glasses) \cite{itoh_towards_2021}, designed for hands-free usage and maintaining situational awareness \cite{orlosky_managing_2014, zhao_headsup_2023, janaka_visual_2022}. There remains a void in crafting interfaces tailored for OST-HMDs suited to daily multitasking, with current messaging apps for OST-HMDs (like Vuzix Blade's WeChat\footnote{\url{https://apps.vuzix.com/app/wechat}}) primarily being derivatives of mobile phone apps (A notable exception is Google Glass XE (2013-2017) \cite{google_glass_voice_2013, google_google_2023} which is discussed in Appendix~\ref{sec:appendix:compare_google_glass}).

The inclination to communicate while multitasking is evident in messaging app usage \cite{curry_most_2022}, fostering closeness and support \cite{cho_i_2020, grinter_wan2tlk_2003}. Mobile phones, while supporting multitasking, can be hazardous in situations needing acute awareness, such as walking \cite{hashish_texting_2017, sullman_cant_2021}. OST-HMDs appear promising due to their hands-free nature and enhanced situational awareness \cite{orlosky_managing_2014, lucero_notifeye_2014}. Voice input stands out as a feasible hands-free technique for OST-HMDs, as other methods like head and gaze inputs might be less accurate or result in ergonomic strain \cite{lee_interaction_2018, cohen_role_1995, revilla_testing_2020}.

Consequently, we present \GlassTexting{} \cite{janaka_glassmessaging_2023}, a messaging application for OST-HMDs, iteratively designed post examining the prevailing needs, habits, challenges, and constraints users encounter while messaging and multitasking.

\begin{figure*}[hptb]
    \centering \small
    \includegraphics[width=0.9\linewidth]{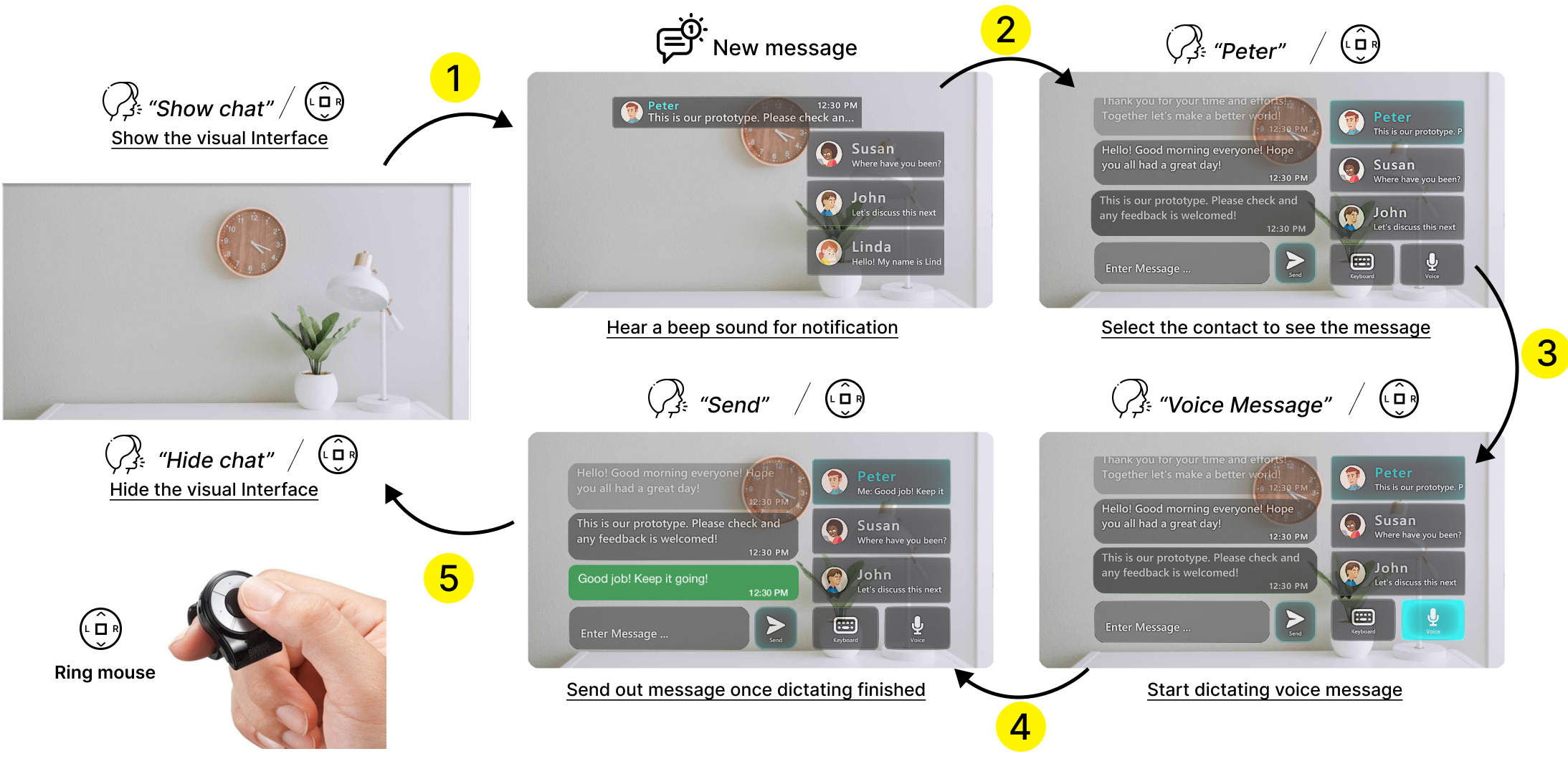}
    \caption{Steps for sending a message after receiving a notification. The user wears the OST-HMD and a ring mouse and sees the environment. The user
    \b{(1) says \voicecommand{show chat}}, the interface is displayed, and a notification including the name of the sender and the sending time appears at the top of the view with a beep sound;
    \b{(2) says the name of the contact (e.g., \voicecommand{Peter})}, and the system automatically navigates to the chat interface of the respective contact;
    \b{(3) says \voicecommand{voice message} and dictates the message via voice}. The system transcribes the user's utterances to text in real-time, displayed in the text entry box. Once the user stops speaking for a measured amount of time (silence gap), dictation turns off automatically;
    \b{(4) says \voicecommand{send}}, and the system sends the message;
    \b{(5) says \voicecommand{hide chat}}, and the full interface is hidden, restoring the full vision of the environment.}
\label{fig:glass_texting:reply_message_steps}
\Description{The figure presents a flowchart consisting of five elements, each depicting a step to follow after receiving a notification. The steps are: (1) The user utters 'show chat', and the chat interface along with a notification appears. (2) The user mentions a contact's name, and the system navigates to the chat interface of that contact. (3) The user speaks a message, and the system transcribes it into text in the text entry box. (4) The user says 'send', and the system sends the message. (5) The user utters 'hide chat', and the full interface is hidden, restoring the complete view of the environment.}
\end{figure*}

\section{System}

After evaluating existing mobile messaging apps (e.g., WhatsApp, Telegram) on OST-HMDs by sideloading, we found that, while their UI was generally intuitive, they were not optimized for OST-HMDs \cite{janaka_glassmessaging_2023}. For example, content often obstructed the view, the color schemes were either too bright or too dark, and some elements were too small. These factors led to usability issues. To cater to hands-busy scenarios, we introduced \textbf{voice dictation} for text entry and \textbf{voice commands} for hands-free UI navigation. We also implemented a \textbf{ring mouse interaction} to allow for faster and more precise scrolling and selection \cite{sapkota_ubiquitous_2021}, while retaining \textbf{mid-air gestures} due to their \quote{intuitive} touch-like content manipulation paradigm.

\subsection{Apparatus}

We selected Microsoft HoloLens 2 (HL2), an OST-HMD with hand-tracking, voice commands, and world-scale positioning (2k resolution, 52° diagonal FoV), to develop \GlassTexting{}, our messaging app designed for OST-HMDs. A wireless ring mouse (Sanwa Supply 400-MA077) facilitated easy directional UI element selection (Figure~\ref{fig:glass_texting:reply_message_steps}). We developed \GlassTexting{} using Unity 3D, Mixed Reality Toolkit (MRTK 2.8), leveraging MRTK's built-in functions for mid-air gestures, voice inputs, virtual keyboard, and content stabilization. To simulate a realistic messaging experience, we implemented a virtual chat server using Python, running on a tablet computer connected to the HL2 via Wi-Fi, enabling bi-directional communication through a socket connection between the client and the server (see \url{https://github.com/NUS-HCILab/GlassMessaging}). 

\subsection{Interface Design}

To enhance learnability and maintain consistency \cite{nielsen_enhancing_1994} with familiar interfaces, we chose to modify the UIs of existing mobile \texting{} apps and tailor them to OST-HMDs, instead of entirely redeveloping them. The final interface is shown in \textbf{Figure~\ref{fig:glass_texting:reply_message_steps}} after two iterations (see details at \cite{janaka_glassmessaging_2023}).

\subsubsection{Visual interface (output)}

The visual interface of \GlassTexting{} (Figure~\ref{fig:glass_texting:reply_message_steps}) consists of four main UI panels, namely, notifications, contacts, chat messages, and voice/keyboard input panels. This allows users to receive notifications, select contacts, and compose/send messages using voice and keyboard input.

\subsubsection{Audio interface (Input-Output)}
As depicted in Figure~\ref{fig:glass_texting:reply_message_steps}, users can interact with \GlassTexting{} via voice commands (Table~\ref{tab:glass_texting:interactions}) to navigate the UI (e.g., \voicecommand{scroll up}, \voicecommand{scroll to top}) and dictate text (using \voicecommand{voice message}). Audio feedback (e.g., beeps) accompanies some input interactions.

When the app is not in dictation mode, voice commands can directly activate various functionalities, such as opening notifications (\voicecommand{open notification}),  selecting contacts (\voicecommand{\NAME{}}), sending the message (\voicecommand{send}), and hiding the interface (\voicecommand{hide chat}).
Voice shortcuts such as \voicecommand{text \NAME{}} are also available, which combine \voicecommand{\NAME{}} and \voicecommand{voice message} for direct text entry. Similarly, the \voicecommand{reply} command opens the notification and begins dictation for a reply immediately.

\subsubsection{Manual-input interface (Input)}
\GlassTexting{} supports two manual input methods: a wearable ring-mouse and mid-air hand gestures as shown in Table~\ref{tab:glass_texting:interactions}.

\textit{Ring mouse}: 
The user can scroll through the contact list using the ring mouse's \buttoncommand{up} and \buttoncommand{down} buttons. The \buttoncommand{right} button toggles between input modalities and selects the send button. The \buttoncommand{center} button activates the selected virtual button and serves as a long-press toggle to hide/reveal the entire interface.

\textit{Mid-air interaction}: 
The visual interface can be interacted with through mid-air gestures. The contact list can be scrolled by swiping, and a contact's chat can be opened by pressing their virtual icon. The input modality is chosen by selecting the corresponding virtual buttons (voice or keyboard). Pressing on a notification opens the chat with the sender.

\section{Evaluation}
To assess the effectiveness of \GlassTexting{}, we compared it to the Telegram application on mobile phones in a controlled study set in daily multitasking situations. Our findings \cite{janaka_glassmessaging_2023} indicate that, even with the present technological constraints of the OST-HMD platform, \GlassTexting{} provided enhanced voice input access and enabled smoother interactions than phones. This resulted in a 33.1\% reduction in response time and a 40.3\% increase in texting speed. These findings underscore the significant potential of OST-HMDs as a meaningful complement to mobile phone-based messaging in multitasking scenarios.

However, there are several challenges to overcome before fully harnessing this platform's potential. For example, the use of \GlassTexting{} resulted in a 2.5\% drop in texting accuracy, especially with complex texts. Moreover, current OST-HMDs have some inherent downsides (e.g., rudimentary hardware capabilities, unfamiliarity, limited interactions \cite{technavio_global_2022, itoh_towards_2021, lee_interaction_2018}) when contrasted with the mature and extensively tested mobile phones currently available.

\section{Conclusion and Future Work}

While multitasking with messaging is a frequent real-life activity, current mobile applications and platforms fall short in providing adequate support. We pinpointed two primary situational impediments (i.e., hands-busy and eyes-busy) arising from existing mobile platforms, which drove us to iteratively develop \href{https://github.com/NUS-HCILab/GlassMessaging}{\GlassTexting{}}, a messaging application tailored for OST-HMDs to address these shortcomings. 
We envision messaging on OST-HMDs as the forthcoming communication frontier, acting as a valuable adjunct to mobile phones during multitasking and driven forward by technological progress. To realize this vision, it is essential to re-conceptualize communication interfaces that align with OST-HMD affordances and to devise strategies to overcome potential situational challenges (e.g., privacy and social concerns with voice).

\begin{acks}
We thank the volunteers who participated in our studies.

This research is supported by the National Research Foundation, Singapore, under its AI Singapore Programme (AISG Award No: AISG2-RP-2020-016). It is also supported in part by the Ministry of Education, Singapore, under its MOE Academic Research Fund Tier 2 programme (MOE-T2EP20221-0010), and by a research grant \#22-5913-A0001 from the Ministry of Education of Singapore. Any opinions, findings and conclusions, or recommendations expressed in this material are those of the author(s) and do not reflect the views of the National Research Foundation or the Ministry of Education, Singapore.
\end{acks}

\bibliographystyle{ACM-Reference-Format}
\bibliography{template.bib}

\appendix

\begin{table*}[hbtp]
\centering
\small
\caption{Supported input interactions of \GlassTexting{}}
\Description{The table presents the supported input interactions of the final version of GlassMessaging. It outlines the different functionalities, such as revealing the interface, hiding the interface, opening chats, activating voice dictation, sending messages, opening and closing the virtual keyboard, navigating between contacts, and initiating voice dictation to specific contacts. Each functionality is listed along with the associated input methods: Mid-air gesture, Ring interaction, and Voice command.}
\begin{tabular}{@{}llll@{}}
\toprule
\textbf{Function} & \textbf{Mid-air gesture} & \textbf{Ring interaction} & \textbf{Voice command} \\ \midrule
Reveal interface &  & Click \buttoncommand{center} button for 1 second & Show chat \\ \midrule
Hide interface &  & Click \buttoncommand{center} button for 1 second & Hide chat \\ \midrule
\begin{tabular}[c]{@{}l@{}}Open the chat related \\ to notification\end{tabular} & Press on notification &  & Open notification \\ \midrule
\begin{tabular}[c]{@{}l@{}}Open the chat with the \\ contact, \NAME{}\end{tabular} & Press on contact \NAME{} & \begin{tabular}[c]{@{}l@{}}Click \buttoncommand{up}/\buttoncommand{down} button to \\ navigate\end{tabular} & \NAME{} \\ \midrule
Activate voice dictation & Press on \buttoncommand{voice} button & \multicolumn{1}{c}{\multirow{3}{*}{\begin{tabular}[c]{@{}c@{}}Click \buttoncommand{right} button to navigate + \\ click \buttoncommand{center} button to activate\end{tabular}}} & Voice message \\ \cmidrule(r){1-2} \cmidrule(l){4-4} 
Send the message & Press on \buttoncommand{send} button & \multicolumn{1}{c}{} & Send \\ \cmidrule(r){1-2} \cmidrule(l){4-4} 
Open the virtual keyboard & Press on \buttoncommand{keyboard} button & \multicolumn{1}{c}{} & Open keyboard \\ \midrule
Close the virtual keyboard & Press anywhere on the interface & Click any button & Close keyboard \\ \midrule
Go to the topmost contact & Scroll up using the finger & Click and hold the \buttoncommand{up} button & Scroll to the top \\ \midrule
Go to closest top contact & Scroll up using the finger & Click \buttoncommand{up} button & Scroll up \\ \midrule
Go to closest bottom contact & Scroll down using the finger & Click \buttoncommand{down} button & Scroll down \\ \midrule
\begin{tabular}[c]{@{}l@{}}Start voice dictating to \\ message received contact\end{tabular} &  &  & \begin{tabular}[c]{@{}l@{}}Reply \\ (Open notification + Voice message)\end{tabular} \\ \midrule
\begin{tabular}[c]{@{}l@{}}Start voice dictation to \\ the contact, \NAME{}\end{tabular} &  &  & \begin{tabular}[c]{@{}l@{}}Text \NAME{} \\ (\NAME{} + Voice message)\end{tabular} \\ \bottomrule
\end{tabular}
\label{tab:glass_texting:interactions}
\end{table*}

\section{\GlassTexting{} vs. Google Glass XE}
\label{sec:appendix:compare_google_glass} 

Google Glass XE (2013-2017) \cite{google_glass_voice_2013, google_google_2023, wikipedia_google_2023}, a discontinued product, supported heads-up messaging. Here, we distinguish between our application and Google Glass XE, showcasing our contributions from both practical and academic perspectives.

\subsection{Google Glass XE (\GoogleGlass{}) interface}

\GoogleGlass{} incorporated a default set of voice action commands for messaging \cite{google_glass_voice_2013}. Its lightweight and seamless design combined voice, head gestures, and touch gestures for inputs and an OST-HMD for output. To activate voice commands or send messages, users would utter \quote{OK Glass} and \quote{Send a message to}, followed by the contact's name and message content. Users would respond to a message by saying \quote{Reply} followed by their message content. Hence, \GoogleGlass{} provided an efficient method for sending and replying to \textit{individual} messages.

\subsection{Comparison}

\begin{table*}[hptb]
\centering
\small
\caption{A comparison between \GlassTexting{} and Google Glass XE. LoS stands for Line of Sight, and FoV represents Field of View. Note: This list is not exhaustive and based on public online resources \cite{google_google_2023, wikipedia_google_2023, niora_google_2023}, as Google Glass XE has been discontinued since 2017.}
\begin{tabular}{@{}lll@{}}
\toprule
\textbf{Features} & \b{\GlassTexting{}} with HL2 &\b{Google Glass XE (\GoogleGlass{})} \\ \midrule
Interactions & Voice, Ring-mouse, Mid-air & Voice, Touchpad (on the right temple), Head gestures \\ \midrule
Text entry & Voice, Mid-air keyboard & Voice \\ \midrule
\multirow{2}{*}{Display} & Binocular, Higher-Resolution (2048x1080 px per eye) & Monocular, Lower-resolution (640x360 px, right eye) \\ 
& Larger-FoV (30\textdegree{} horizontal) & Smaller-FoV (13\textdegree{} horizontal) \\ \midrule
Chat history & Shows last 3 messages and 3 contacts & Shows last message and last contact \\ \midrule
Chat position & LoS (middle-center) & \multirow{3}{*}{Above LoS (top-center), Manual switching between each UI} \\ 
Notification position & Above LoS (top-center) &  \\ 
Contact position & Right of LoS (middle-right) &  \\ \midrule
UI opacity & Increased for new messages & Fixed \\ \midrule
UI access & On-demand (using voice or ring-mouse) & On-demand (by looking up or using voice) \\ 
\bottomrule
\end{tabular}
\label{tab:compare_google_glass}
\end{table*}

Table~\ref{tab:compare_google_glass} depicts that both \GlassTexting{} (\GlassMessaging{}) and \GoogleGlass{} utilize voice input for text entry and navigation. Our study \cite{janaka_glassmessaging_2023} validates voice input as an efficient tool, aligning with \GoogleGlass{}'s design. However, speech recognition affects the accuracy of \GlassMessaging{}, a challenge possibly shared by \GoogleGlass{} users. \GoogleGlass{}, while catering to immediate messaging requirements, had difficulty managing intricate conversations. In contrast, \GlassMessaging{} upholds modern standards, emphasizing context via features like full chat history and unread indicators. The display location differs too: \GoogleGlass{} showcased content above the line-of-sight, demanding attention shifts, while \GlassMessaging{}, leveraging advancements, positions content within the line-of-sight, employing opacity adjustments for awareness.

Interaction-wise, \GoogleGlass{} relied on head-gestures, while \GlassMessaging{} introduced a gamut of methods like ring and mid-air gestures, providing flexibility for multitasking. Ultimately, while both platforms serve heads-up messaging, their design nuances cater to different generational hardware and user demands. \GoogleGlass{}, tailored for earlier-generation glasses, prioritized real-time singular messages. \GlassMessaging{}, on the other hand, leverages advanced OST-HMDs, managing both immediate and layered messaging. A fusion of their strengths, such as integrating \GoogleGlass{}'s head-gesture system into \GlassMessaging{}, could potentially amplify user experience, especially in intricate messaging scenarios.

\end{document}